\documentclass{aastex631}
\usepackage{xcolor}
\usepackage{soul}

\begin{document}

\title{Search for neutrino signals correlated with LHAASO diffuse Galactic emission}

\author[0000-0002-4300-5130]{Wenlian Li$^{\dagger}$}
\affiliation{Tsung-Dao Lee Institute, Shanghai Jiao Tong University, 201210 Shanghai, China}
\affiliation{Key Laboratory for Particle Astrophysics and Cosmology (MoE), Shanghai Key Laboratory for Particle Physics and Cosmology, 200240 Shanghai, China}

\author[0000-0001-8510-2513]{Tian-Qi Huang$^{\dagger}$}
\affiliation{Key Laboratory of Particle Astrophysics and Experimental Physics Division and Computing Center, Institute of High Energy Physics, Chinese Academy of Sciences, 100049 Beijing, China}
\affiliation{TIANFU Cosmic Ray Research Center, Chengdu, Sichuan, China}

\author[0000-0003-1639-8829]{Donglian Xu}
\affiliation{Tsung-Dao Lee Institute, Shanghai Jiao Tong University, 201210 Shanghai, China}
\affiliation{Key Laboratory for Particle Astrophysics and Cosmology (MoE), Shanghai Key Laboratory for Particle Physics and Cosmology, 200240 Shanghai, China}
\affiliation{School of Physics and Astronomy, Shanghai Jiao Tong University, 200240 Shanghai, China}

\author[0000-0002-5963-4281]{Huihai He}
\affiliation{Key Laboratory of Particle Astrophysics and Experimental Physics Division and Computing Center, Institute of High Energy Physics, Chinese Academy of Sciences, 100049 Beijing, China}
\affiliation{TIANFU Cosmic Ray Research Center, Chengdu, Sichuan, China}
\affiliation{University of Chinese Academy of Sciences, 100049 Beijing, China}


\correspondingauthor{D.L. Xu}
\email{donglianxu@sjtu.edu.cn}

\def\thefootnote{$\dagger$}\footnotetext{These authors contributed equally to this work}\def\thefootnote{\arabic{footnote}}










\begin{abstract}


The diffuse Galactic $\gamma$-ray emission originates from the interactions between cosmic rays and the interstellar medium or radiation fields within our Galaxy, where the production of neutrinos is also anticipated. Recently, the Large High Altitude Air Shower Observatory (LHAASO) reported measurements of diffuse $\gamma$-rays from the Galactic plane with energies ranging from sub-TeV to 1 PeV. Using publicly available 7 years of IceCube track data with the full detector, we conduct a template search using the $\gamma$-ray flux map observed by LHAASO-KM2A as the neutrino emission template and perform a scan search of the Galactic plane. In the template search, a mild excess of neutrinos is observed in the Galactic plane with a pretrial \textcolor{black}{(posttrial)} significance of $1.9\sigma$ \textcolor{black}{($1.1\sigma$)}. The measured muon neutrino intensity at 25 TeV is $4.73^{+2.53}_{-2.51}\times10^{-14}\,{\rm TeV^{-1}\,cm^{-2}\,s^{-1}\,sr^{-1}}$, consistent with the expected neutrino flux assuming that all the diffuse Galactic $\gamma$-rays originate from hadronic interactions. In the Galactic plane scan search, the most significant location is found at $l=63.57^{\circ}$ and $b=0.93^{\circ}$ with a pretrial (posttrial) significance of $4.6\sigma$ ($1.8\sigma$).

\end{abstract}



\section{Introduction} \label{sec:intro}


The origin and acceleration mechanism of the high-energy cosmic rays remain a puzzle since the discovery of cosmic rays in 1912 \citep{hess1912observations}. During the propagation within the Galactic plane, cosmic rays interact with the dense interstellar medium through deep inelastic scattering, producing multiple secondary particles. Among these particles, neutral pions ($\pi^{0}$) decay into $\gamma$-rays, while charged pions ($\pi^{\pm}$) decay into neutrinos. 
\textcolor{black}{These electrically neutral particles can point back to their origin when detected on Earth, making the Galactic plane an ideal ``cloud chamber'' to trace these cosmic rays.}


The diffuse Galactic $\gamma$-ray emission (DGE) from the Galactic plane has been measured by many experiments from sub-GeV to PeV \citep{Fermi-LAT:2012edv, 2008ApJ...688.1078A, 2014PhRvD..90l2007A, 2015ApJ...806...20B, TibetASgamma:2021tpz, 2022icrc.confE.835., LHAASO:2023gne}. The DGE primarily originates from the decay of neutral pions, bremsstrahlung radiation, and the inverse Compton scattering. 
\textcolor{black}{The hadronic ($\pi^{0}$-decay) component is predicted to dominate at higher energies by some theoretical models \citep[e.g.,][]{Gaggero:2015xza,Lipari:2018gzn}, where a corresponding diffuse Galactic neutrino flux is expected to associate with these $\pi^{0}$-decay $\gamma$-rays.}


Galactic neutrino searches have been conducted in the past decade \textcolor{black}{\st{for}} many times \citep[e.g.,][]{IceCube:2017trr, ANTARES:2017nlh, ANTARES:2018nyb, IceCube:2019lzm}. Most searches are based on templates which provide prior distributions of neutrino signals in spatial and energy domains. 
IceCube identified neutrino emission from the Galactic plane at the $4.5\sigma$ level of significance using the cascade events \textcolor{black}{collected} from 2011 to 2021 \citep{Abbasi:2023bvn}, based on the Fermi Large Area Telescope (LAT) $\pi^{0}$ \citep{Fermi-LAT:2012edv}, \textcolor{black}{KRA$^5_\gamma$, and KRA$^{50}_\gamma$} models \citep{Gaggero:2015xza}.
ANTARES also observed the diffuse neutrino emission from the Milky Way with a posttrial p-value equivalent to $1.7\sigma$ for the KRA$^5_\gamma$ model \citep{Cartraud:2023usm}, and found a mild excess over the background in the Galactic Ridge at $\sim96\%$ confidence level (C.L.) for the isotropic flux template \citep{ANTARES:2022izu}, using both track and cascade events. 
Using the template built on the TeV $\gamma$-ray observation by HAWC, IceCube constrained that $\pi^{0}$-decay $\gamma$-rays contribute no more than 80\% of $\gamma$-rays from the Northern Galactic plane at the 90\% C.L., using the eight-year muon-track data \citep{Kheirandish:2019bke}. 
However, $\gamma$-rays sources are not masked in this template.
With the high-energy starting events released by IceCube, an anisotropic component of neutrino flux was found in the direction of low Galactic latitudes \textcolor{black}{at $\sim 3\sigma$} \citep{Neronov:2015osa} and $4.1\sigma$ \citep{2022ApJ...940L..41K}, \textcolor{black}{both} are template-independent. The latter study also estimated the flux of Galactic neutrinos but didn't consider the instrument response in detail. 
In this study, we use the IceCube muon-track data from 2011 to 2018 to conduct a template search for diffuse Galactic neutrinos originating from the Galactic plane and try to constrain the hadronic contribution to the observed DGE. 
The precise measurement of DGE from 10 TeV to 1 PeV by LHAASO provides a brand new template for diffuse Galactic neutrino emission. \textcolor{black}{It is also} the first measurement of DGE from the outer Galactic plane \citep{LHAASO:2023gne}.
The $\gamma$-ray and neutrino dataset, and the LHAASO and IceCube detectors, are presented in Section \ref{sec_2_detector_and_datasets}. Section \ref{sec_3_analysis_method} presents the analysis methods. Discussions and conclusions are presented in Sections \ref{sec_4_discussion} and \ref{sec_5_conclusion}.

\section{IceCube \& LHAASO Detectors and Datasets}\label{sec_2_detector_and_datasets}

\subsection{IceCube and Neutrino Dataset}

\textcolor{black}{The IceCube Neutrino Observatory, located at the South Pole, is a Cherenkov detector array targeting for astrophysical neutrinos with energies above (a few) TeV \citep{IceCube:2016zyt}. 
Charge current interactions between high-energy muon neutrinos and the ice medium can produce muons, generating a track-like morphology that recorded by the digital optical modules. These events are referred to as track events. 
The track data has the best sensitivity to the Northern Hemisphere due to the Earth's shielding effect \textcolor{black}{to atmospheric muons.}}
\textcolor{black}{The muon-track data includes three parts: the experimental data events, the instrument response functions, and the detector uptime \citep{2021arXiv210109836I}. Only the events observed by the full 86-string detector are used in these analyses.} \textcolor{black}{These events were recorded between May 2011 and July 2018, with a total livetime of 2531.9 days and a total of 897,406 events.}

\subsection{LHAASO and $\gamma$-Ray Dataset}

\textcolor{black}{LHAASO is a mega-scale composite instrument designed to study $\gamma$-rays and cosmic rays \citep{2019arXiv190502773C}. \textcolor{black}{LHAASO ($100.01^{\circ}\mathrm{E}$, $29.35^{\circ}\mathrm{N}$) has a wide field of view and covers the sky from a declination of $-21^{\circ}$ to $79^{\circ}$.} The Kilometer Square Array (KM2A) of LHAASO is optimized for detecting $\gamma$-rays with energies ranging from 10 TeV to a few PeV,} \textcolor{black}{while the Water Cherenkov Detector Array (WCDA) of LHAASO operates in lower energy ranges from 100 GeV to $20$ TeV \citep{hehh:2018}.}


Recently, LHAASO-KM2A reported measurements of diffuse $\gamma$-rays from the Galactic plane with energies ranging from 10 TeV to 1 PeV \citep{LHAASO:2023gne}. The observation of the DGE by LHAASO was conducted in two regions: the inner Galaxy region ($15^{\circ} < l < 125^{\circ}$, $|b| < 5^{\circ}$) and the outer Galaxy region ($125^{\circ} < l < 235^{\circ}$, $|b| < 5^{\circ}$). 
\textcolor{black}{All known point-like and extended sources detected by KM2A \citep{LHAASO:2023rpg}, as well as those from TeVCat \citep{2008ICRC....3.1341W}, were masked.}
The diffuse $\gamma$-rays follow a power-law spectrum ${\rm d}N/{\rm d}E \propto E^{-2.99}$ for both the inner and outer regions, \textcolor{black}{collectively referred to as the LHAASO region hereafter.} 
\textcolor{black}{The corresponding flux map and significance map were released with a pixel size of $\Delta l = 2^{\circ}$ and $\Delta b = 1^{\circ}$.}
\textcolor{black}{LHAASO-WCDA also observed the diffuse Galactic $\gamma$-rays below 20 TeV, following \textcolor{black}{an} $E^{-2.64}$ spectrum in the inner region and \textcolor{black}{an} $E^{-2.60}$ spectrum in the outer region \citep{Li:2023dpg}.}

\section{Analysis}\label{sec_3_analysis_method}

\subsection{Template Search}
%



Since the emission from the Galactic plane is quite extensive and exhibits diffuse morphology, the unbinned maximum likelihood commonly used in neutrino point-source searches \citep{BRAUN2008299} is not suitable for this analysis. 
Instead, we use the ps-template likelihood, as illustrated in \citep{IceCube:2017trr}, to search for neutrino emission from the Galactic plane. 
The signal-subtracted template likelihood \citep{Pinat:2017wxs, Pinat:2017ldg} is defined as:
\begin{equation}
L(n_s,\gamma)  = \prod_{i=1}^{N}\Big(\frac{n_s}{N}S_i(\mathbf{x}_i ,\sigma_i,E_i;\gamma)+\widetilde{D}_i(sin\delta_i,E_i)-\frac{n_s}{N}\widetilde{S}_i(sin\delta_i,E_i)\Big)
\end{equation}\label{eq_template_likelihood}
where $n_s$ is number of signal events, \textcolor{black}{$\gamma$ represents spectral index}, $N$ is the total number of events, $S_i$ is the signal probability density function (PDF) \textcolor{black}{of the i-th event}, $\widetilde{D}_i$ is the scrambled background PDF estimated from data, and $\widetilde{S}_i$ is the scrambled signal PDF. Each PDF comprises a spatial term and an energy term. The details of the construction of the likelihood follow the methods described in \citep{Li:2024gnb}. 

In the template search, we only fit the number of signal events while the neutrino spectral shape is fixed. 
The test statistic is defined as the log-likelihood ratio
$TS = 2 {\rm ln} \left[L(\hat{n}_s)/L(n_s=0) \right]$ to derive the significance and upper limits. 
\textcolor{black}{The probability distribution for $TS$ is approximately a $\chi^2_{1}$ distribution for the background hypothesis \citep{d543aecb-cd73-36d5-9101-f08a74f8e8c6}.}
\textcolor{black}{The $1\sigma$ uncertainties are given by ${\rm ln}L(n_{\pm 1\sigma})={\rm ln}L(\hat{n}_s)-1/2$, and the 90\% C.L. upper limits are given by ${\rm ln}L(n_{90})={\rm ln}L(\hat{n}_s)-1.64/2$.}

\subsubsection{$\gamma$-Ray Flux Templates}

\textcolor{black}{\textcolor{black}{The flux template uses the $\gamma$-ray flux map observed by LHAASO-KM2A} as the weighting for neutrino emission, assuming that all the $\gamma$-rays originate from neutral pion decay in p-p interactions.} \textcolor{black}{The observed $\gamma$-ray flux ($\phi_{\gamma}$) and single-flavor neutrino flux ($\phi_{\nu}$) are related as}
\begin{equation}
\phi_{\gamma}(E_{\gamma}) = \frac{1}{2}\phi_{\nu}(E_{\nu})e^{-\tau_{\gamma\gamma}}, 
\end{equation}
\textcolor{black}{where the $\gamma$-ray energy $E_{\gamma}=2E_{\nu}$ and the $\gamma$-ray attenuation $e^{-\tau_{\gamma\gamma}}$ is negligible in the template search (see \autoref{appendix:absorption}).}
\textcolor{black}{Thus, the neutrino spectrum breaks at 5 TeV, following the KM2A spectral index above this threshold and the WCDA spectral index below it.}

\begin{figure}[htbp]
    \centering
    \includegraphics[width=0.95\textwidth]{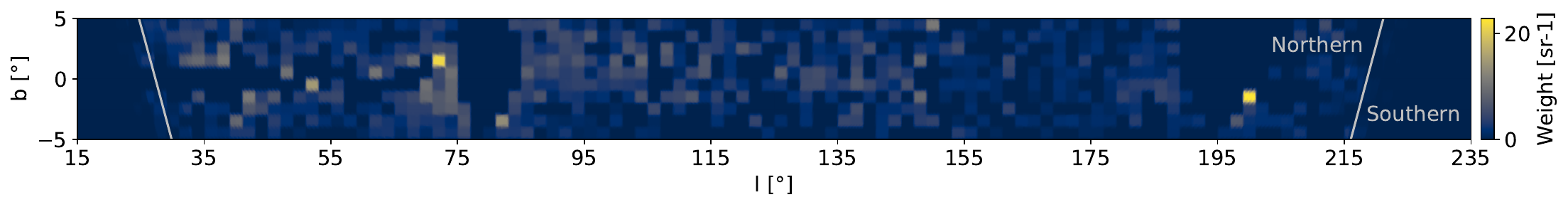}
    \caption{\textcolor{black}{The $\gamma$-ray flux map ($0.5\sigma$) convolved with IceCube detector acceptance, shown in Galactic coordinates. The silver line indicates the horizon between the Northern and Southern skies, with a boundary at decl. $=-5^{\circ}$.}}
    \label{gamma_template_acceptance} 
\end{figure}

To mitigate the effect of background fluctuations \textcolor{black}{on the $\gamma$-ray flux map}, we apply significance cuts to the flux map, testing various significance thresholds, including $0.5\sigma$, $1.0\sigma$, $1.5\sigma$, and $2.0\sigma$. 
\textcolor{black}{This means only pixels with significance $> 0.5\sigma$, $1.0\sigma$, $1.5\sigma$, or $2.0\sigma$ in the flux map are retained for the spatial template.} \textcolor{black}{These areas are referred to as LHAASO $>n\sigma$ regions ($n=0.5,\,1.0,\,1.5,\,2.0$).}
\textcolor{black}{The neutrino emission template derived from the $\gamma$-ray flux map with $0.5\sigma$ significance cut is shown in Figure \ref{gamma_template_acceptance}, convolved with the IceCube detector acceptance.}

\subsubsection{Other Templates}

\textcolor{black}{
We also test five other templates.
The uniform template assumes a uniform distribution of neutrinos from the Galactic plane ($15^{\circ}<l<235^{\circ}$, $|b|<5^{\circ}$).
The gas template ($15^{\circ}<l<235^{\circ}$, $|b|<5^{\circ}$) assumes that the neutrino flux is proportional to the gas column density traced by the PLANCK dust opacity map \citep{Planck:2016frx}, under the assumption of a uniform ratio between dust opacity and gas column density. 
The three all-sky templates are based on the Galactic diffuse neutrino emission models, referred to as $\pi^0$ \citep{Fermi-LAT:2012edv}, $\rm KRA_{\gamma}^5$, and $\rm KRA_{\gamma}^{50}$ \citep{Gaggero:2015xza}.
Furthermore, we test these templates in the same region as LHAASO's analysis for a more direct comparison with the $\gamma$-ray flux templates.
}

\textcolor{black}{The uniform and gas templates assume the same neutrino spectrum as the $\gamma$-ray flux templates. For the $\pi^0$ and $\rm KRA_{\gamma}$ templates, the spectra are fixed to $E^{-2.7}$ and the model-averaged spectrum over the sky, respectively.}

\subsection{Galactic Plane Scan Search}

\textcolor{black}{We further scan the Galactic plane 
to study the spatial correlation of neutrino hot spots with the gas distribution and the $\gamma$-ray distribution.}
\textcolor{black}{The Galactic plane ($15^{\circ} < l < 235^{\circ}$, $|b| < 5^{\circ}$) is binned into equal-sized pixels ($\sim 0.2^{\circ}\times 0.2^{\circ}$) with a HEALPix resolution parameter Nside=256 \citep{Gorski:2004by}.}
At \textcolor{black}{the center} of each pixel, we conduct a point-source search assuming a power-law spectrum.
The point-source likelihood function is defined as
\begin{equation}
L(n_s,\gamma)  = \prod_{i=1}^{N}\Big(\frac{n_s}{N}S_i(\mathbf{x}_i ,\sigma_i,E_i;\mathbf{x}_s,\gamma)+(1-\frac{n_s}{N}){B_i}({\rm sin}\delta_i,E_i)\Big),
\end{equation}
where $n_s$ and $\gamma$ are \textcolor{black}{both free parameters} for each pixel. The signal PDF $S_i$ and the background PDF $B_i$ are constructed \textcolor{black}{following the methods summarized in \citep{IceCube:2016tpw, Huang:2021hjc}.} \textcolor{black}{We use the maximum-likelihood ratio method, which compares the hypothesis of the point-like signals plus background to the background-only null hypothesis (atmospheric background and isotropic astrophysical neutrinos).}

\textcolor{black}{We obtain the $TS$ distribution for each pixel in the Galactic plane under the background hypothesis through pseudo-experiments. The pseudo-experiment data samples are obtained by randomizing the R.A. of events in IceCube data. The observed $TS$ values are then converted to p-values using these $TS$ distributions. 
\textcolor{black}{For $TS < 5$, the p-value is estimated directly from pseudo-experiments. For $TS \geq 5$, a truncated gamma function is applied to describe the tail of the $TS$ distribution, following the method summarized in \citep{2022Sci...378..538I}.}
}

\section{Results and Discussion}\label{sec_4_discussion}
\subsection{Template Search Results}

The results for the template searches are summarized in Table~\ref{tab_1} and Table~\ref{tab_2}. Although some excesses from the \textcolor{black}{Galactic plane} are observed, the results are not statistically significant. 
\textcolor{black}{
For the $\gamma$-ray flux map with a $0.5\sigma$ significance threshold, we obtained the smallest pretrial p-value of $0.029$, corresponding to a coincidence significance of $1.9\sigma$. When using the gas template, a pretrial p-value of $0.067$ $(1.5\sigma)$ is obtained.
\textcolor{black}{For the $\pi^0$ and $\rm KRA_{\gamma}^5$ model templates, the pretrial p-values of \textcolor{black}{0.070 $(1.5\sigma)$} and \textcolor{black}{0.19 $(0.9\sigma)$} are obtained.}
The results obtained from the $\gamma$-ray flux map are more significant than those from the uniform ($1.0\sigma$), gas ($1.0\sigma$), and $\pi^0$  \textcolor{black}{($1.2\sigma$)} templates with 
the same mask region \textcolor{black}{as that applied in LHAASO's analysis.}
\textcolor{black}{For each template, the $90\%$ C.L. upper limit on the muon neutrino flux is set. The best-fit fluxes are also listed, with the $1\sigma$ uncertainty given only for those templates with a pretrial significance larger than $1.0\sigma$.} \textcolor{black}{With different significance cuts on the $\gamma$-ray flux map}, the resulting upper limits are $\sim 3$ times higher than the theoretical prediction assuming all the $\gamma$-rays are of hadronic origin (Table \ref{tab_1}).}

\textcolor{black}{The posttrial significance for the $\gamma$-ray flux map ($0.5\sigma$) template is obtained through pseudo-experiments. The posttrial p-value is determined as the fraction of pseudo-experiments in which at least one of the considered templates is more significant than $1.9\sigma$. The posttrial p-value is reduced to 0.14 ($1.1\sigma$) after correcting for trial factors across all templates in \autoref{tab_1} and \autoref{tab_2}.}



\begin{table}[htbp]
    \centering
    \begin{tabular}{cccccc}
        \hline
        \hline
        Spatial Template & $\hat{n}_s$ & Pretrial p-value \textcolor{black}{($\sigma_{\rm pre}$)} & Best-fit flux & Upper Limit $\phi_{90\%}$  & \textcolor{black}{$\phi_{90\%}/ \phi_{\nu}$}  \\
        \hline
        Flux map (0.5$\sigma$) & $311.4$ & $0.029 (1.9\sigma)$ & $1.78^{+0.95}_{-0.94} \times 10^{-14}$ & $3.00\times 10^{-14}$ & $2.9$  \\
        Flux map (1.0$\sigma$) & $278.8$ & $0.036 (1.8\sigma)$ & $1.56^{+0.88}_{-0.87} \times 10^{-14}$ & $2.68\times 10^{-14}$ & $2.9$  \\
        Flux map (1.5$\sigma$) & $244.5$ & $0.040 (1.8\sigma)$ & $1.34^{+0.78}_{-0.77} \times 10^{-14}$ & $2.35\times 10^{-14}$ & $3.0$  \\
        Flux map (2.0$\sigma$) & $182.5$ & $0.064 (1.5\sigma)$ & $0.98^{+0.66}_{-0.65} \times 10^{-14}$ & $1.82\times 10^{-14}$ & $3.2$  \\
        \hline
    \end{tabular}
    \caption{\textcolor{black}{Results of template searches using \textcolor{black}{$\gamma$-ray} flux map templates with different significance thresholds. For each template, the best-fit signal event number $\hat{n}_s$ is provided, along with the pretrial p-value, best-fit neutrino flux, 90\% C.L. upper limit on neutrino flux, and the ratio between the upper limit and the theoretically predicted neutrino flux. \textcolor{black}{The best-fit flux and the upper limit flux are given at the $E_{\nu}=25\,{\rm TeV}$ in the unit of $\mathrm{TeV}^{-1}\mathrm{cm}^{-2}\mathrm{s}^{-1}$.} The theoretical neutrino flux is predicted under the assumption that diffuse Galactic $\gamma$-rays are entirely from p-p interactions.}}
    %
    \label{tab_1}
\end{table}

\begin{table}[htbp]
    \centering
    \begin{tabular}{ccccc}
        \hline
        \hline
        Spatial Template & $\hat{n}_s$ & Pretrial p-value \textcolor{black}{($\sigma_{\rm pre}$)} & Best-fit flux & Upper Limit $\phi_{90\%}$  \\
        \hline
        Uniform (with mask) & $208.9$ & $0.15 (1.0\sigma)$ & $1.21^{+1.16}_{-1.16} \times 10^{-14}$ & $2.71 \times 10^{-14}$ \\
        Gas (with mask) & $181.2$ & $0.16 (1.0\sigma)$ & $1.03 \times 10^{-14}$ & $2.36\times 10^{-14}$  \\
        $\pi^0$ (with mask) & \textcolor{black}{$210.3$} & \textcolor{black}{$0.12 (1.2\sigma)$} & \textcolor{black}{$1.61^{+1.36}_{-1.35} \times 10^{-14}$} & \textcolor{black}{$3.35\times 10^{-14}$} \\
        $\rm KRA_{\gamma}^5$ (with mask) & \textcolor{black}{$154.7$} & \textcolor{black}{$0.13 (1.1\sigma)$} & \textcolor{black}{$0.22_{-0.20}^{+0.20}\times$MF} & \textcolor{black}{$0.48\times$MF} \\
        $\rm KRA_{\gamma}^{50}$ (with mask) & \textcolor{black}{$124.3$} & \textcolor{black}{$0.15 (1.0\sigma)$} & \textcolor{black}{$0.16_{-0.16}^{+0.16}\times$MF} & \textcolor{black}{$0.37\times$MF} \\
        \hline
        Uniform & $336.8$ & $0.069 (1.5\sigma)$ & $1.92^{+1.30}_{-1.29} \times 10^{-14}$ & $3.58\times 10^{-14}$  \\
        Gas & $276.5$ & $0.067 (1.5\sigma)$ & $1.62^{+1.09}_{-1.08} \times 10^{-14}$ & $3.01\times 10^{-14}$  \\
        $\pi^0$ & \textcolor{black}{$553.7$} & \textcolor{black}{$0.070 (1.5\sigma)$} & \textcolor{black}{$7.79^{+5.35}_{-5.30} \times 10^{-14}$} & \textcolor{black}{$1.46\times 10^{-13}$} \\
        $\rm KRA_{\gamma}^5$ & \textcolor{black}{$163.5$} & \textcolor{black}{$0.19 (0.9\sigma)$} & \textcolor{black}{$0.59\times$MF} & \textcolor{black}{$1.48\times$MF} \\
        $\rm KRA_{\gamma}^{50}$ & \textcolor{black}{$123.5$} & \textcolor{black}{$0.23 (0.8\sigma)$} & \textcolor{black}{$0.39\times$MF} & \textcolor{black}{$1.09\times$MF} \\
        \hline
    \end{tabular}
    \caption{\textcolor{black}{Results of template searches. Similar to Table \ref{tab_1}, but for the uniform, gas, $\pi^0$, $\rm KRA_{\gamma}^5$, and $\rm KRA_{\gamma}^{50}$ templates. The first five rows show the templates masked to \textcolor{black}{the LHAASO region}. The last five rows show the templates without any mask. The best-fit flux and 90\% C.L. upper limit flux are presented in units of the model flux (MF) for the $\rm KRA_{\gamma}$ templates and are given at $E_{\nu}=25\,{\rm TeV}$ in units of $\mathrm{TeV}^{-1}\mathrm{cm}^{-2}\mathrm{s}^{-1}$ for other templates.}}
    \label{tab_2}
\end{table}

\begin{figure}[htbp]
    \centering
    \includegraphics[width=0.65\textwidth]{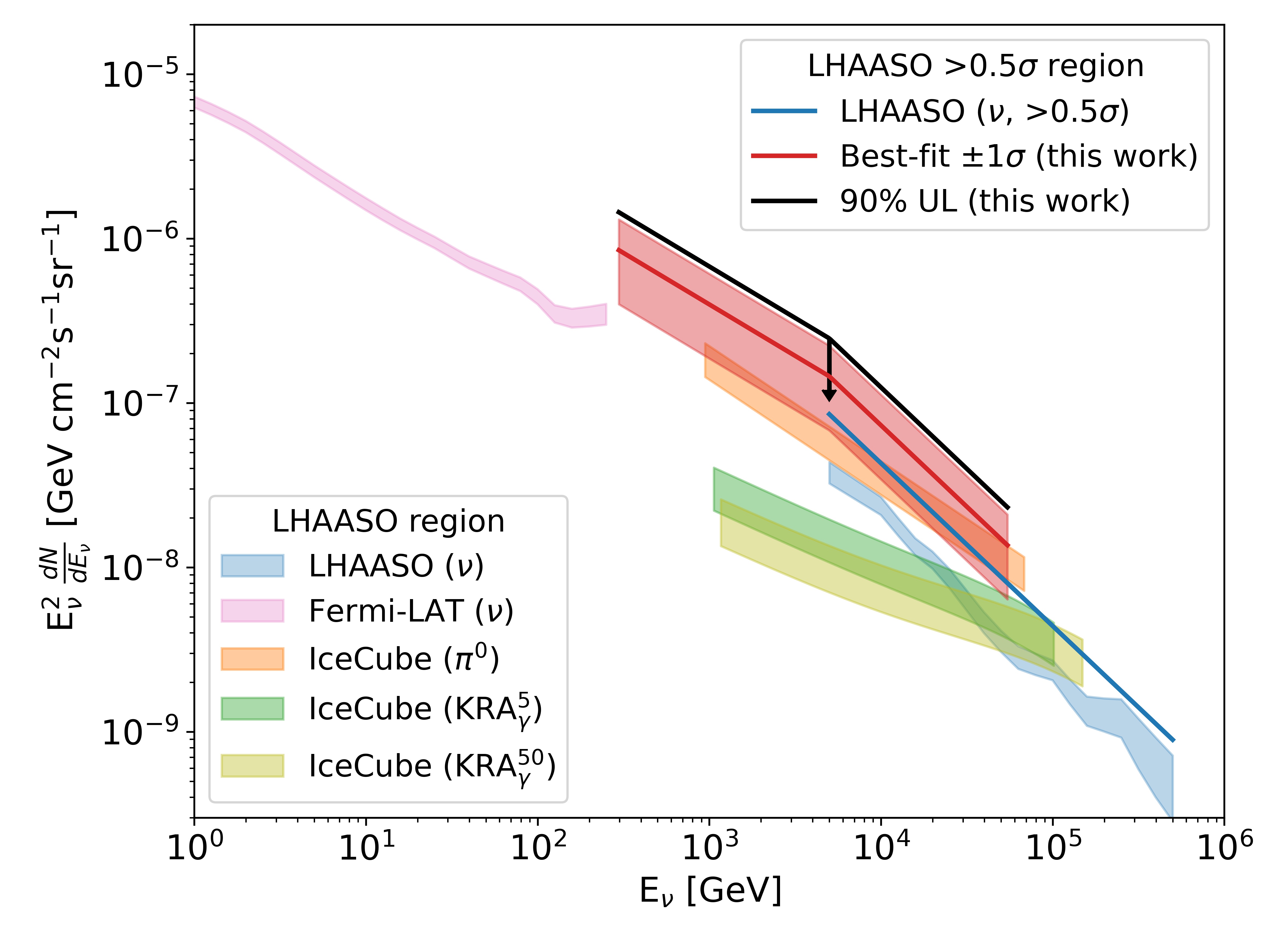}
    \caption{The muon neutrino intensity as a function of neutrino energy $E_{\nu}$. The best-fit \textcolor{black}{(90\% C.L. upper limit)} flux obtained \textcolor{black}{using the $\gamma$-ray flux map with $> 0.5\sigma$ detection} is shown as a red \textcolor{black}{(black)} solid line, \textcolor{black}{averaged over the LHAASO $>0.5\sigma$ region.} The red shaded region indicates the $1\sigma$ uncertainty and extends over the energy range contributing 90\% significance (for details, see Appendix \ref{appendix:energy_range}). \textcolor{black}{The theoretically predicted muon neutrino flux, derived from the LHAASO DGE observations \citep{LHAASO:2023gne} and Fermi-LAT DGE observations from \textcolor{black}{the LHAASO region} \citep{Zhang:2023ajh} assuming hadronuclear interactions, are shown as the blue and pink shaded areas, respectively.} After applying a 0.5$\sigma$ significance cut to the LHAASO $\gamma$-ray flux map, the corresponding intensity is shown as a blue solid line, which is approximately twice the intensity without significance cuts \textcolor{black}{(blue shaded area). The diffuse neutrino flux measured by IceCube \citep{Abbasi:2023bvn} using $\pi^0$ (orange) and $\rm KRA_{\gamma}$ templates (green and olive) are also shown. These fluxes are scaled by multiplying the sky-integrated flux with the relative template contribution from the \textcolor{black}{LHAASO region} and averaged over this region.}
    }
    \label{galactic_plane_upper_limit} 
\end{figure}

\textcolor{black}{Figure~\ref{galactic_plane_upper_limit} shows the best-fit flux with $1\sigma$ uncertainty (red shaded area) obtained in the template searches using the $\gamma$-ray flux map with $> 0.5\sigma$ detection}, compared with the theoretically predicted muon neutrino flux \textcolor{black}{assuming all observed DGE \textcolor{black}{in the LHAASO $>0.5\sigma$ region} are of hadronic origin (blue solid line).}
The theoretically predicted flux (blue solid line) in the same sky region is within the best-fit flux $1\sigma$ uncertainty \textcolor{black}{($4.73^{+2.53}_{-2.51}\times10^{-14}\,{\rm TeV^{-1}\,cm^{-2}\,s^{-1}\,sr^{-1}}$ at 25 TeV)} and is lower than the upper limit flux. 
Therefore, our findings \textcolor{black}{are consistent with} the pure hadronic origin of the DGE observed by LHAASO. 

Figure~\ref{galactic_plane_upper_limit} also shows the neutrino flux measured by IceCube \citep{Abbasi:2023bvn} and the theoretically predicted neutrino flux based on LHAASO DGE observations \citep{LHAASO:2023gne} and Fermi-LAT DGE observations \citep{Zhang:2023ajh}. 
\textcolor{black}{The latter only provides the flux measurements in the LHAASO region, so we average the neutrino flux in these measurements within this region.}
The average neutrino flux in the LHAASO $>0.5\sigma$ region is higher than the average flux in the LHAASO region.
This phenomenon is likely a result of the applying the significance cut to the $\gamma$-ray flux map that masks the sky region with low neutrino emission.
\textcolor{black}{Additionally, although both regions mask the same sources, the bin sizes are different.}
The deviation from the measurements of IceCube at the lower energy range is probably due to the different energy spectra and the different data samples. The best-fit Galactic diffuse neutrino flux using 12.3 years of track data in the Northern sky is a factor of $\sim 2$ higher than the all-sky cascade analysis for $\rm KRA_{\gamma}$ templates \citep{IceCube:2023hou}.

\textcolor{black}{While we use track-like events in this study due to their good sensitivity in the Northern sky and large statistics, cascades will facilitate searches for signals from very extended regions like the Galactic plane due to their lower backgrounds. Analyses using combined tracks and cascades may offer enhanced sensitivity \textcolor{black}{\citep{IceCube:2023gtp}} and could be investigated in the future.}


\textcolor{black}{
In the case of templates, more neutrino emission templates can be tested in the future. 
Firstly, the central 90\% energy range is from \textcolor{black}{0.3 TeV to 54.6 TeV} (Figure \ref{galactic_plane_upper_limit}), and some models predict that hadronic emission is more dominant in the lower energy range \citep{Marinos:2022tdj}. In the future, we can use neutrino emission templates based on observations by \textcolor{black}{LHAASO-WCDA}.
Secondly, Galactic plane neutrinos tend to concentrate towards the Galactic Center region, where there are hot spots \citep{Abbasi:2023bvn}. Future large zenith angle diffuse $\gamma$-ray observations by LHAASO towards the GC region, along with observations from future ground-level particle detectors in the Southern sky, such as the Southern Wide-field Gamma-ray Observatory (SWGO) \citep{Conceicao:2023tfb}, will enhance the search for Galactic neutrinos.
}

\subsection{Galactic Plane Scan Search Results}

\begin{figure}[htbp]
   \centering
   \includegraphics[width=0.99\textwidth]{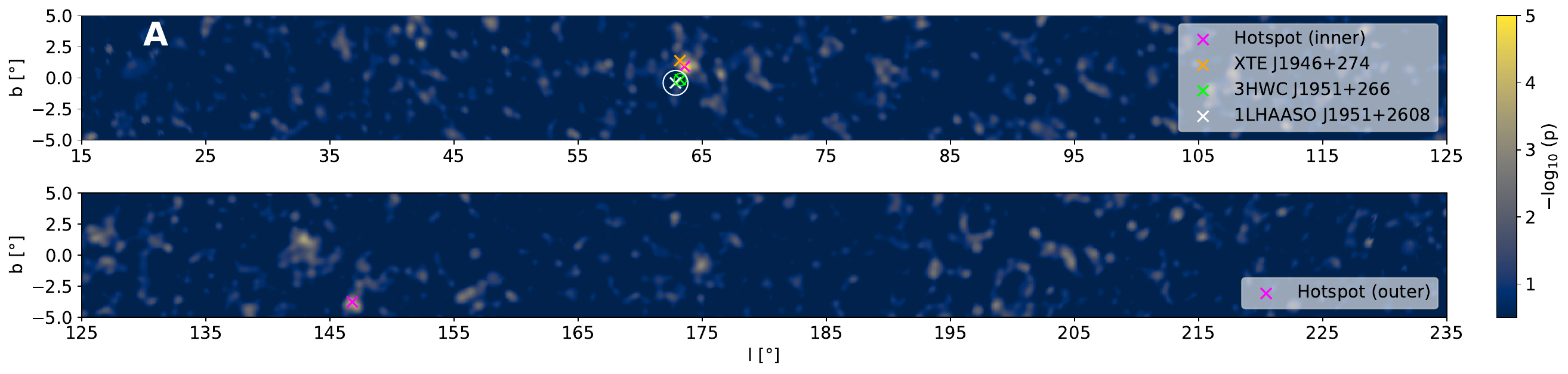}
   \includegraphics[width=0.99\textwidth]{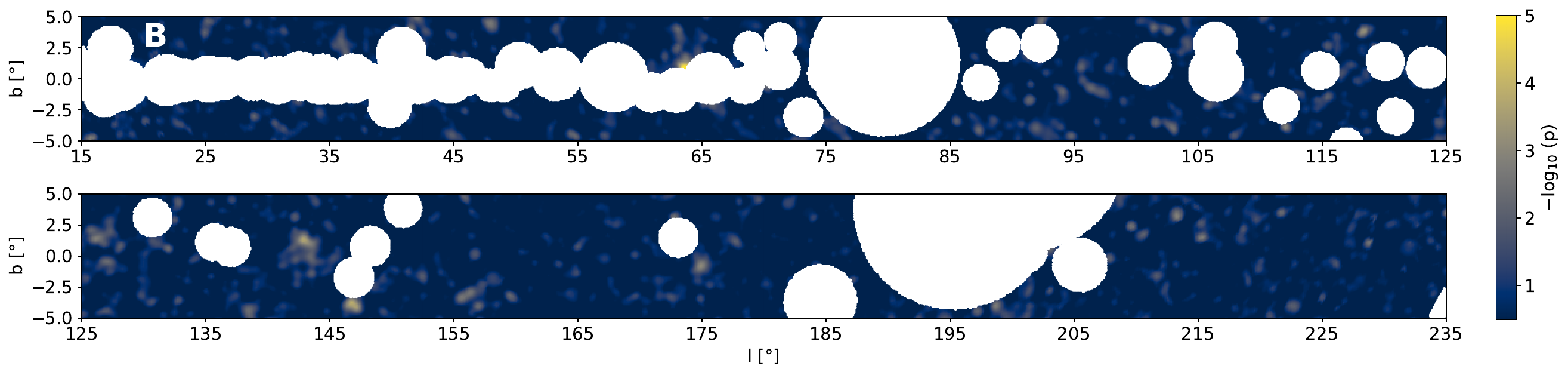}
   \includegraphics[width=0.99\textwidth]{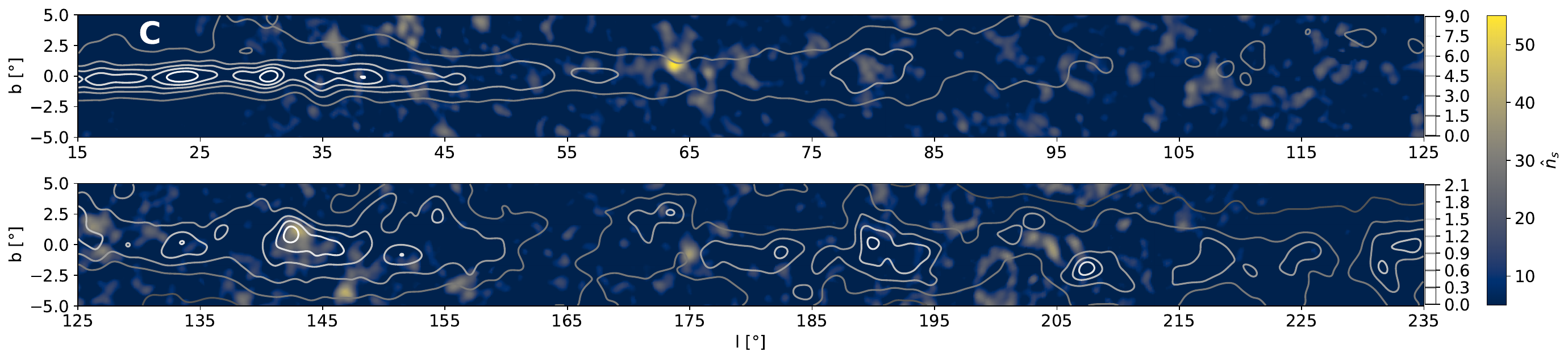}
   \caption{\textcolor{black}{Sky map of the Galactic plane scan for point sources, shown in Galactic coordinates. (A) The color scale indicates the logarithm of the pretrial p-value. The most significant spots in the inner ($l = 63.57^{\circ}$, $b=0.93^{\circ}$) and outer ($l = 146.81^{\circ}$, $b=-3.77^{\circ}$) Galactic plane are indicated by magenta crosses. The nearest X-ray binary and TeV $\gamma$-ray sources to the inner hotspot are also shown as crosses, with the circles indicating the extension of the $\gamma$-ray sources. (B) The color scale is the same as that in (A), with the white region indicating the mask region applied in LHAASO's analysis. (C) The color scale indicates the best-fit $\hat{n}_s$ obtained from the maximum likelihood analysis. Contours indicate the spatial distribution of the gas template traced by the PLANK dust opacity map, smoothed with a Gaussian kernel of $\sigma=0.5^{\circ}$ for comparison with the neutrino excess ($\hat{n}_s$) map. The color scale of the contours indicates the dust column density times $10^4$.}}
   \label{galactic_plane_scan} 
\end{figure}

\textcolor{black}{The sky map of the Galactic plane scan is shown in Figure \ref{galactic_plane_scan}. Figure~\ref{galactic_plane_scan}A and B show the pretrial p-value map, while Figure~\ref{galactic_plane_scan}C shows the neutrino excess ($\hat{n}_s$) map.} 
The hottest spots in the inner and outer \textcolor{black}{Galactic plane} are indicated as magenta crosses in Figure~\ref{galactic_plane_scan}A. For the inner \textcolor{black}{Galactic plane}, the hotspot is located at \textcolor{black}{$l = 63.57^{\circ}$ and $b=0.93^{\circ}$ with a pretrial p-value of $1.9 \times 10^{-6}$ $(4.6\sigma)$ and a posttrial p-value of 0.018 ($2.1\sigma$).} 
\textcolor{black}{The posttrial p-value increases to 0.038 ($1.8\sigma$) when considering the look-elsewhere effect across both the inner and outer regions.}
\textcolor{black}{The best-fit parameters at this spot are $\hat{n}_s=56.2$ and \textcolor{black}{$\hat{\gamma}=3.00$}, corresponding to a muon neutrino flux of $2.52\times10^{-11}\,{\rm TeV^{-1}\,cm^{-1}\,s^{-1}}$ at 1 TeV.} 
For the outer \textcolor{black}{Galactic plane}, the hotspot is found at 
\textcolor{black}{$l = 146.81^{\circ}$ and $b=-3.77^{\circ}$ with a pretrial p-value of \textcolor{black}{$1.1 \times 10^{-4}$ $(3.7\sigma)$} and a posttrial p-value of 0.69.} \textcolor{black}{The best-fit parameters are $\hat{n}_s=43.3$ and $\hat{\gamma}=3.10$.}
The masked point-like and extended sources as in LHAASO's analysis are indicated as white circle in Figure~\ref{galactic_plane_scan}B. 
The contours of the gas distribution as traced by the PLANK dust opacity map \citep{Planck:2016frx} is shown in Figure~\ref{galactic_plane_scan}C, which was plotted on the neutrino excess map. 

\textcolor{black}{
A Be/X-ray binary, XTE J1946+274 ($l = 63.21^{\circ}$, $b=1.40^{\circ}$) \citep{1998IAUC.7014....1S}, is located $0.6^{\circ}$ away from the inner hotspot. This binary exhibited X-ray outburst during various periods \citep{Chandra:2023gad}.
It was found with a pretrial p-value of 0.026 (0.009) in the periodic (flare) search for neutrino emission \citep{IceCube:2022jpz}. \textcolor{black}{We also conduct a time-integrated search assuming a point source. The pretrial p-value is \textcolor{black}{$2.6\times 10^{-3}$ ($2.8\sigma$)}, and the best-fit parameters are $\hat{n}{_s}=36.1$ and $\hat{\gamma}=2.98$.}} 

\textcolor{black}{IceCube has searched for extended sources in the Galactic plane using muon track data from 2011 to 2020 \citep{IceCube:2023ujd}. 
In the catalog search, the most significant location is centered on 3HWC J1951+266 with a radius of $1.7^{\circ}$ at $2.6\sigma$ post-trials. The inner hotspot is correlated with this extended region, being only $1.1^{\circ}$ from 3HWC J1951+266. 1LHAASO J1951+2608 ($l = 62.85^{\circ}$, $b = -0.39^{\circ}$) is associated with 3HWC J1951+266 and is the closest source in the 1LHAASO catalog to the inner hotspot, with an angular separation of $1.5^{\circ}$ \citep{LHAASO:2023rpg}.
The hottest spots in their Galactic plane scan, assuming source extensions of $0.5^{\circ}$ to $2^{\circ}$, are only $0.1^{\circ}$ to $0.3^{\circ}$ away from our inner hotspot.}

\textcolor{black}{There are some hotspots in the region where no $\gamma$-ray sources have been observed (Figure~\ref{galactic_plane_scan}B). }
\textcolor{black}{In the outer Galactic plane, a cluster of neutrino hotspots around $\sim 138^{\circ}<l<142^{\circ}, |b|<2.5^{\circ}$ is spatially associated with a large gas clump (Figure~\ref{galactic_plane_scan}C).}
There are also some hotspots in the $\gamma$-ray observation in this region \citep{LHAASO:2023gne}. 
\textcolor{black}{Since neutrinos and diffuse $\gamma$-rays from neutral pion decay mainly come from the hadronic interactions of cosmic rays and interstellar gas, gas clumpy regions will have higher probabilities of producing neutrinos and $\gamma$-rays. 
However, this cluster is not significant, with the best-fit spectral index $\hat{\gamma}$ of around 4. 
\textcolor{black}{A similar cluster is also found in the same region of IceCube Northern-sky scan \citep{2022Sci...378..538I}, but it is not observed in the Galactic plane scan for extended sources \citep{IceCube:2023ujd}. In the IceCube all-sky scan using cascade events \citep{Abbasi:2023bvn}, a cluster is found  \textcolor{black}{near} the position ($l=145^{\circ}$, $b=-10^{\circ}$), that is more than $10^{\circ}$ away. This excess is likely \textcolor{black}{due to statistical} fluctuation.}}

\section{Conclusion and Outlook}\label{sec_5_conclusion}


\textcolor{black}{In this study, we conduct two searches for neutrino signals from the Galactic plane using 7 years of publicly released track data observed by the IceCube full detector: a template search and a scan search. Mild excesses of neutrinos are found in both searches. In the template search, the most significant neutrino excess is found in the flux template based on the LHAASO-KM2A $\gamma$-ray flux map with a significance threshold of $0.5\sigma$, yielding a pretrial p-value of 0.029 ($1.9\sigma$)\textcolor{black}{, which is reduced to $1.1\sigma$ when the trial factor is accounted for.} \textcolor{black}{The measured neutrino intensity, $4.73^{+2.53}_{-2.51}\times10^{-14}\,{\rm TeV^{-1}\,cm^{-2}\,s^{-1}\,sr^{-1}}$ at 25 TeV, is consistent with the prediction assuming that all the $\gamma$-rays have a hadronic origin.} In the Galactic plane scan search, the most significant point is found at $l=63.57^{\circ}$ and $b=0.93^{\circ}$ with a pretrial p-value of $1.9\times10^{-6}$ ($4.6\sigma$) and a posttrial 
\textcolor{black}{p-value of 0.038 ($1.8\sigma$).}
}

\textcolor{black}{In the future, neutrinos from the Galactic plane can be further investigated using additional templates based on $\gamma$-ray observations, such as the diffuse Galactic $\gamma$-rays observed by LHAASO-WCDA and future observations of the Galactic Center region. Moreover, combined analyses using more data, including both tracks and cascades observed by ANTARES, IceCube, KM3NeT-ARCA \citep{2016JPhG...43h4001A}, Baikal-GVD \citep{2021chep.confE.606S}, and the future neutrino telescopes such as IceCube-Gen2 \citep{2021JPhG...48f0501A}, P-ONE \citep{2020NatAs...4..913A}, TRIDENT \citep{2023NatAs...7.1497Y}, and HUNT \citep{Huang:2023R8}, along with additional templates, will elucidate the propagation of cosmic rays in the Galaxy.}



\section*{Acknowledgments} \label{sec:ackno}

We thank the LHAASO collaboration for helpful discussions during the preparation of this work. D.L.X. and W.L.L. acknowledge the National Natural Science Foundation of China (NSFC) grant (No. 12175137) on “Exploring the Extreme Universe with Neutrinos and multi-messengers” and the Double First Class start-up fund provided by Shanghai Jiao Tong University. T.-Q.H. acknowledges the support of the Special Research Assistant Funding Project of the Chinese Academy of Sciences. H.H.H. acknowledges the support of NSFC grant (No. 12105294) and the Innovative Project of Institute of High Energy Physics (No. E45454U210).

\appendix

\section{$\gamma$-Ray Absorption}\label{appendix:absorption}

\begin{figure}[htbp]
    \centering
    \includegraphics[width=0.65\textwidth]{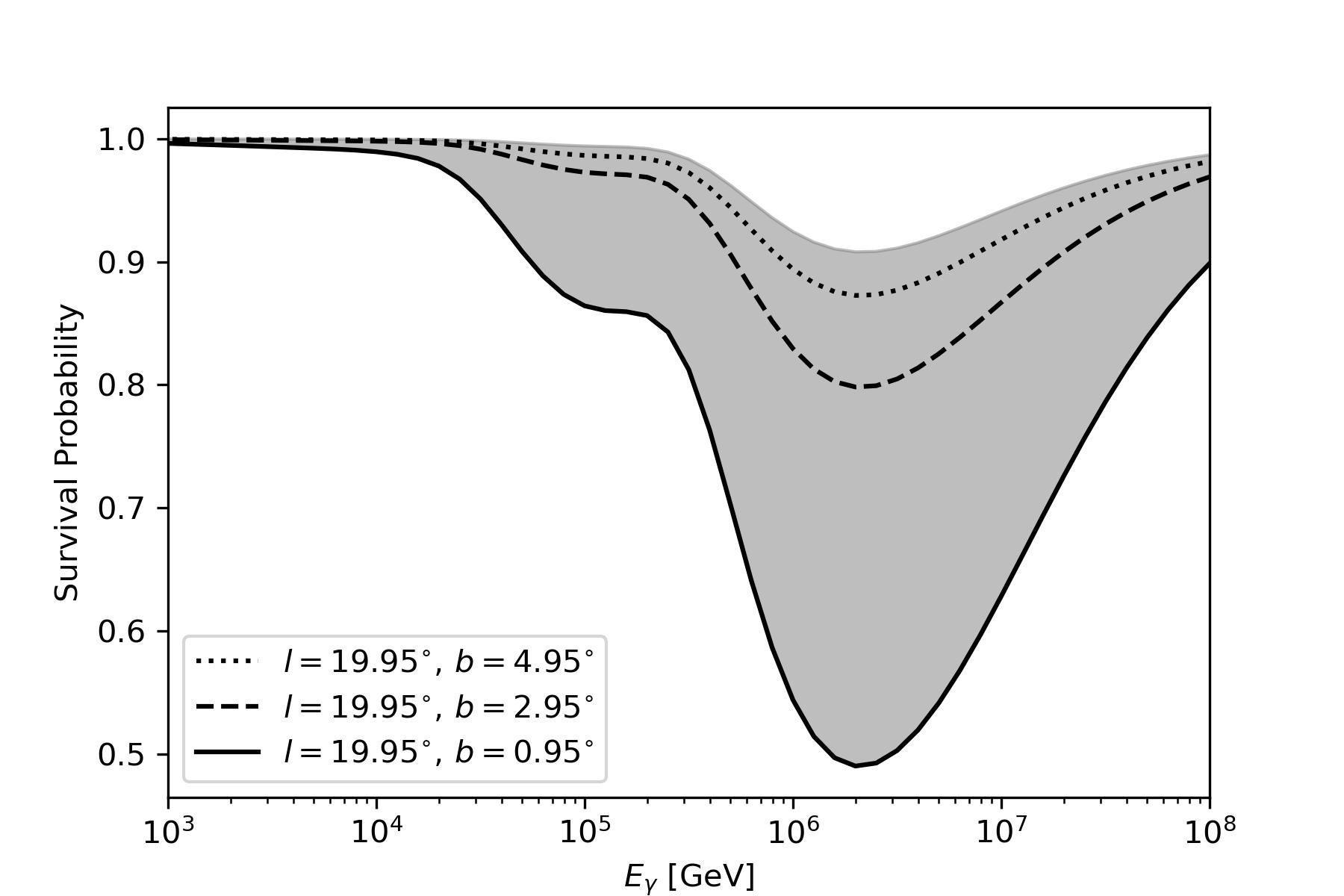}
    \caption{Survival probability of diffuse Galactic $\gamma$-rays from different directions in the \textcolor{black}{LHAASO region}.}
    \label{survival probability} 
\end{figure}

We consider the absorption to diffuse $\gamma$-rays from the Galactic plane. 
The diffuse $\gamma$-ray emission density is proportional to the product of cosmic ray and gas densities. 
The cosmic ray density is assumed to follow the spatial distribution of supernova remnants \citep{2015MNRAS.454.1517G, 2020A&A...643A.137S}, while the gas density is assumed to follow the HI gas model \citep{2003PASJ...55..191N, 2017JCAP...02..015E}, following the method in \citep{2023ApJ...957L...6F}. 
The $\gamma$-ray absorption is dominated by the interstellar radiation field \citep{2017MNRAS.470.2539P} and the cosmic microwave background through pair production ($\rm \gamma+\gamma\rightarrow e^{+}+e^{-}$). 
In the \textcolor{black}{LHAASO region} (see \autoref{galactic_plane_scan}B), the survival probabilities of diffuse Galactic $\gamma$-rays from different directions fall within the shadow region of Figure \ref{survival probability}. 

Neutrinos above 5 TeV will deviate from the $E^{-2.99}$ spectrum if $\gamma$-ray absorption is considered. However, the absorption has little effect on our results. For the $\gamma$-ray flux map ($0.5\sigma$), in the case of the most extreme absorption \textcolor{black}{(solid line in \autoref{survival probability}), the best-fit muon neutrino flux at 25 TeV is $1.80^{+0.98}_{-0.96}\times10^{-14}\,{\rm TeV^{-1}\,cm^{-2}\,s^{-1}}$ with a pretrial p-value of 0.030, and the central 90\% energy range is from 0.3 to 57.7 TeV.}

\section{Central Energy Range}\label{appendix:energy_range}

The test statistic values contributed by signal events can be expressed as
\begin{equation}
    TS_{s} = 2 \sum_{i} \omega_{i} {\rm ln}\bigg[\frac{\hat{n}_s}{N}\bigg( \frac{S_{i}}{\widetilde{D}_{i}}-\frac{\widetilde{S}_{i}}{\widetilde{D}_{i}} \bigg)+1 \bigg]=\sum_{i}\omega_{i}TS_i,
\end{equation}
where the weight term $\omega_i\propto S_i/\widetilde{D}_i$ is the probability of the ith event being signal. The weight terms satisfy the condition $\sum\omega_i=\hat{n}_s$.
The neutrino energy distribution of $TS_{s}$ follows
\begin{equation}
    \frac{{\rm d} TS_{s}}{{\rm d} E_{\nu}} = \sum_{i}\omega_{i}TS_i P_s(E_{\nu}|E_i)
\end{equation}
where
$P_s(E_{\nu}|E_{i})$ is the probability density distribution of neutrino energy $E_{\nu}$ for signal events with a reconstructed energy of $E_i$. The central 90\% energy range is defined as the neutrino energy range in which astrophysical signals contributing the central 90\% of $TS_{s}$. The central 90\% energy range for the $\gamma$-ray flux map (0.5$\sigma$) template is from \textcolor{black}{0.3 to 54.6 TeV.}



\bibliography{sample631}{}
\bibliographystyle{aasjournal}



\end{document}